\documentclass[prl,twocolumn,showpacs,preprintnumbers,amsmath,amssymb]{revtex4}
\usepackage{simplewick}
\newcommand {\con}[3]{\contraction[0.5ex]{}{#1}{#2}{#3}\nomathglue{#1#2#3}}
\usepackage{graphicx}
\usepackage{dcolumn}
\usepackage{bm}
\newcommand \be{\begin{eqnarray}}
\newcommand \ee{\end{eqnarray}}
\newcommand \ba{\begin{align}}
\newcommand \eea{\end{align}}

\newcommand {\f}[1]{(\ref{#1})}
\newcommand {\z}[1]{\cite{#1}}
\newcommand \s{{\backslash\hspace{-1ex}i
}}

\begin{document}
           \csname @twocolumnfalse\endcsname
\title{Equivalence of channel-corrected T-matrix and anomalous propagator approach}
\author{K. Morawetz$^{1,2}$}
\affiliation{$^1$M\"unster University of Applied Science,
Stegerwaldstrasse 39, 48565 Steinfurt, Germany}
\affiliation{$^2$International Institute of Physics (IIP),
Universidade Federal do Rio grande do Norte - UFRN, Brazil
}

\begin{abstract}
Any many-body approximation corrected for unphysical repeated collisions in a given condensation channel is shown to provide the same set of equations as they appear by using anomalous propagators. The ad-hoc assumption in the latter theory about non-conservation of particle numbers can be released. In this way the widespread used anomalous propagator approach is given another physical interpretation. A generalized Soven equation follows which improves a chosen approximation in the same way as the coherent potential approximation (CPA) improves the averaged T-matrix for impurity scattering. 
\end{abstract}
\maketitle

Superconducting and Bose-Einstein condensation phenomena belong to one of the most exciting macroscopic effects based on microscopic quantum physics. The theoretical description of both phenomena is one of the major activity in theoretical many-body physics. Superconductivity is based on the pairing of two fermions which form a condensate while bosons provides a one-particle Bose-Einstein condensate. Both phenomena are characterized by possessing a singular channel in which the effect appears.

The self-consistent multiple scattering T-matrix in a many-body surrounding diverges near the critical temperature of the onset of the symmetry-broken phase, may it be pairing condensation for fermions or Bose-Einstein condensation for bosons. Though describing correctly the onset of pairing, the T-matrix does not provide the gap equation. This changes if an asymmetric breaking of the selfconsistency in the T-matrix is used, such that one of the two internal propagators is used self-consistently and the other non-selfconsistently. Then the gap equation appears as the pole of the T-matrix.
This was first observed by Kadanoff and Martin \cite{KM61} and used later on \cite{MBL99,HCCL07,L08,SLMM09} and is called Prange paradox \cite{Prange60}. It has remained puzzling since a seemingly worse approximation leads to better results. Recently it turned out that the repeated collisions \cite{L08} with the same particle are responsible for this artifact. Repeated collisions of two particles in the same state are unphysical since the particles move apart from each other after the collision. If these repeated collisions with the same state are removed from the T-matrix the correct gap equation appears and the condensate can be described without asymmetrical ad-hoc assumptions about selfconsistency. 
The advantage of eliminating only the contributions of single channels as proposed in Refs.~\cite{L08} and \cite{SLMM09} is that the formation of pairs and their condensation can be described within the same approximation. This has also resulted in the description of different phases in interacting Bose systems \cite{MML09}.

On the other hand there exist a well established theory to describe systems
with condensates in terms of anomalous functions, for review see
\cite{SG98}. Let us consider bosonic particles which can form a condensate
either bosons or paired fermions. If the system processes a macroscopic number $N_0$ of such particles
in the condensate represented by channel $i$, the expectation value of the
creation operator $a_0^+$ of that state is very huge and to a good accuracy
the creation operators $[a_0,a_0^+]\approx 0$ commute with each other as well
as with all other states \z{SG98}. The number of particles is considered as
non-constant if the condensate is thought as a reservoir, since scattering off
and on the condensate may create/destroy pairs. Therefore the anomalous
Green function $G_{12}=\frac 1 i \langle T a_{-k} a_k\rangle$ are nonzero
besides the normal Green function $G_{11}=\frac 1 i \langle T a_{k}
a_k^+\rangle$ describing the simultaneous excitation of a pair. This ad-hoc
assumption leads then to the description of the condensate and the gap
equation for pairing. Please note that the nature of the condensate remains
quite different whether it is composed of bosons or paired fermions \cite{Y62}.

The question is now, how the two approaches above are related. In the first theory correcting the T-matrix we consider only microscopic correlations while the same result is obtained by the second approach where one assumes ad-hoc from the beginning anomalous functions. In the second approach the symmetry of the theory concerning particle conservation is obviously broken while in the first approach it remains conserving. So it seems to be worth to understand the relation between both approaches. 

The aim of the present paper is to show that indeed the first approach leads to the same structure of equations like the second one yielding expressions for the anomalous functions without assuming them. Indeed it will be shown that the theory is somehow overdetermined by the second approach in that one can work with half the number of equations if following the first procedure. In view of this, the virtue of the first approach consists in giving the anomalous function assumption a further microscopic meaning since they can be derived from theory.

Let us therefore shortly sketch the structure of the first approach. We split
the selfenergy into different channels, $\Sigma=\sum_j \Sigma_j$, where we
assume for simplicity that we have only one singular channel $i$ where the condensate appears. The unphysical multiple scatterings with the same channel are concerning a single channel and vanish in the thermodynamical limit. Therefore this deficiency does not matter in normal matter. If we have a singular channel due to the condensate, however, this correction becomes essential.  We have to subtract this process, i.e. we define the subtracted propagator
\be
G_\s=G-G_\s\Sigma_iG
\label{1}
\ee 
or $G^{-1}=G^{-1}_\s-\Sigma_{i}$. Using the standard Dyson equation $G_0^{-1}=G^{-1}+\Sigma$ we obtain the relation
\be
G_\s=G_0+G_0(\Sigma-\Sigma_i) G_\s
\label{2}
\ee
which shows that in this propagator the own selfenergy channel is subtracted, $\Sigma'=\Sigma-\Sigma_i$.
Now we consider a general T-matrix which represents the selfenergy as
$\Sigma'=\sum\limits_{j\ne i} T_j\bar G$ where the channel T-matrix $T_j$ as
two-particle function is closed by an backward propagator $\bar G$. In the
singular channel we subtract the repeated interaction within this channel. This is achieved by closing with the subtracted propagator $\Sigma_i=T_i\bar G_\s$. Now we can rewrite the Dyson equation as
\be 
G^{-1}&=&G_0^{-1}-\Sigma
=G_0^{-1}-\Sigma'-\Sigma_i
\nonumber\\
&=&G_0^{-1}-\Sigma'-T_i\bar G_\s
\nonumber\\
&=&G_0^{-1}-\Sigma'-T_i\left (\bar G_0^{-1}-\bar \Sigma'\right )^{-1}
\label{gm}
\ee
where in the last step we have used \f{2}. Finally we rewrite \f{gm} to obtain
the full propagator in momentum-energy (Matsubara) representation $p=({\bf p},\omega_n)$
\be
G(p)={\bar G_0^{-1}(p)-\bar \Sigma'(p)\over
[G_0^{-1}(p)-\Sigma'(p)][\bar G_0^{-1}(p)-\bar \Sigma'(p)]-T_i(p)}.
\label{10}
\ee
Remembering $\Sigma'=\Sigma-T_i\bar G_\s$ leads immediately back to the Dyson equation $G=G_0/(1-\Sigma G_0)$. Therefore it is an exact rewriting so far.

Now we take into account the explicit form of the free propagator $G_0^{-1}=\omega-\epsilon_k$ and $\bar G_0^{-1}=-\omega-\epsilon_{-k}$ and call the ``proper'' selfenergy
\be
\Sigma_{11}(p)\equiv \Sigma'(p).
\ee
Further, we observe that the T-matrix in the singular channel is separable
\cite{BSI74,BSI75,GZh99} and can be written $T_i(p)=\mp \Delta(p) \bar \Delta(p)$
for bosons/fermions respectively. Now we can define the ``anomalous'' selfenergy as
\be
\Sigma_{12}(p)\equiv \Delta(p)
\ee
such that the propagator \f{10} takes the form
\be
G_{11}={\omega+\epsilon_{-k}+\bar \Sigma_{11} \over (\omega+\epsilon_{-k}+\bar \Sigma_{11})(\omega-\epsilon_k-\Sigma_{11})\pm\Sigma_{12}^2}.
\ee
We add now an auxiliary quantity called ``anomalous'' Green function
\be
G_{12}\equiv {\mp\Sigma_{12} \over (\omega+\epsilon_{-k}+\bar \Sigma_{11})(\omega-\epsilon_k-\Sigma_{11})\pm\Sigma_{12}^2}
\label{aux}
\ee
and can write in such a way the two equations in matrix form
\be
{\bf G=G^0+G^0 \Sigma G}
\ee
with
\ba
{\bf G}\!=\!\left (
\!\!\begin{array}{cc}
G_{11} &G_{12}\\ \bar G_{12} &\pm\bar G_{11}
\end{array}
\!\!\right )\!, 
\,
{\bf G_0}\!=\!\left (
\!\!\begin{array}{cc}
G_0 &0\\ 0 &\pm \bar G_0
\end{array}
\!\!\right )\!,
\,
{\bf \Sigma}\!=\!\left (
\!\!\begin{array}{cc}
\Sigma_{11} &\Sigma_{12}\\ \bar \Sigma_{12} &\pm \bar \Sigma_{11}
\end{array}
\!\!\right )
\nonumber\\
\end{align}
for bosons/fermions respectively.
These are exactly the equations for anomalous propagators derived first by
Beliaev \z{B58} for bosons. For fermions these are the Nambu-Gorkov equations \z{Gor58}.

In other words separating a singular channel from the selfenergy avoiding
repeated collision within this channel leads immediately to propagators which
have the Beliaev form for sosons or the Nambu-Gorkov form for fermions. We see
that adding the auxiliary quantity \f{aux} is not necessary. All information
we have derived without this quantity and it was added here simply to show the
same structure of theory appears as provided by the approaches with anomalous
functions. In this sense the theory of anomalous functions is overdetermined.
We should note, however, that the anomalous propagator $G_{12}$ describes the order parameter. This anomalous propagator appears as a result of the theory here and is not assumed from the beginning as done usually.

Now that we have clarified that the anomalous propagator is an exact rewriting of the Dyson equation if one correct a channel of selfenergy for repeated collisions we might ask what kind of equation such channel-corrected selfenergy obeys. This will lead us to a generalization of the Soven equation \cite{Soven67}. The Soven equation was proposed to describe impurity scattering in terms of an effective medium and resulted in the  coherent potential approximation (CPA) \cite{VKE68,V69,EKL74}. This CPA improves the averaged T-matrix \z{KS54a} with respect to better analytic properties and a wider range of applications. It has turned out that the averaged T-matrix is the uncorrected channel while the CPA is equivalent to the channel-corrected approximation. Here we will present the same idea of channel correction but applied to the two-particle scattering. This will lead to a general Soven equation  which allows to improve a chosen approximation scheme in the same manner as the CPA improves the averaged T-matrix approximation for impurity scattering.

Let s assume in general the defining equation for the channel T-matrix in terms of the potential $V$ and a block $K_j$
\be
T_j=V+VK_jT_j
\label{t}
\ee
covering both the singular channel $j=i$ as well as the normal channels $j\ne i$. In general this equation is a two-particle one which is reduced to the
one-particle selfenergy by closing the upper line with the backward
propagating Green function $\Sigma'=\sum\limits_{j\ne i}\con{T}{_j}{\bar G}$
and for the singular channel $\Sigma_i=\con{T}{_i}{\bar G}_\s$. In the following we will denote explicitly by which function the upper line is closed. All other products are understood as operator products of one-particle functions.
For the above mentioned averaged T-matrix approximation one has $K=G_\s$ and a closing by $\bar G\to c$ in terms of the impurity concentration $c$. In two-particle ladder approximation one would have the form $K_i=G\cdot G_\s$. In the following we consider the general block $K$ such that any more refined approximation can be chosen.

\begin{figure}
\includegraphics[width=8cm]{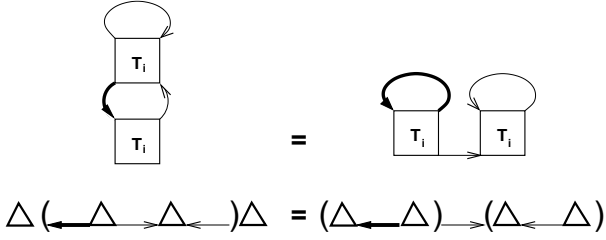}
\caption{Proof of relation (\ref{h3}) for singular channel T-matrices which become separable.\label{fig1}}
\end{figure}

With \f{1} and $ \Sigma_i=\con{T}{_i}{ \bar G_\s}$ we can write
\be
1=(1-D) G ^{-1} G_\s +DG ^{-1}G_\s+G ^{-1}G_\s \con{T}{_i}{ \bar G_\s}G
\label{hh1}
\ee
where we have added and subtracted an operator $D$ which will be determined later by convenience. Now it easy to proof with the help of the separability of the singular channel $T_i=\mp\Delta\Delta$ that the following relation holds
\be
\con{T}{_i}{\bar G}_\s G
=\con{T}{_i}{\bar G} G_\s.
\label{ide}
\ee
Indeed we have with the help of (\ref{1})
\be
\con{T}{_i}{\bar G} G_\s G^{-1}=\con{T}{_i}{\bar G}-\con{T}{_i}{\bar G} G_\s\con{T}{_i}{\bar G}_\s
\label{h1}
\ee
and 
\be
\con{T}{_i}{\bar G}_\s=\con{T}{_i}{\bar G}-\con{T}{_i}{\overline {G_\s \Sigma_i G}}.
\label{h2}
\ee
Since we have the identity \be
\con{T}{_i}{\overline {G_\s \Sigma_i G}}=\con{T}{_i}{\bar G} G_\s\con{T}{_i}{\bar G}_\s
\label{h3}
\ee
as shown in figure \ref{fig1}, the equations (\ref{h1}) and (\ref{h2}) are identical and relation (\ref{ide}) is proved.

Therefore we can write for (\ref{hh1})
\be
1=(1-D) G ^{-1} G_\s +(DG^{-1}+G^{-1}G_\s \con{T}{_i}{ \bar G})G_\s
\label{c1}
\ee
Multiplying \f{1} from the right with $\con{T}{_i}{\bar G_\s}$ 
we can find 
\be
G^{-1}G_\s \con{T}{_i}{\bar G}+\Sigma_i G_\s \con{T}{_i}{\bar G}&=&\con{T}{_i}{\bar G} 
=\con{V}{}{\bar G}+\con{V}{KT_i}{\bar G}
\label{tinv1}
\ee
where we have used \f{t} for the second identity.
For the last term we define now an effective potential $\tilde V_i$
\be
\tilde V_i G_\s \con{T}{_i}{\bar G}\equiv \con{V}{KT_i}{\bar G}
\label{veff}
\ee
with the help of which we can invert \f{tinv1}
\be
G_\s\con{T}{_i}{\bar G}=(G^{-1}+\Sigma_i-\tilde V_i)^{-1}\con{V}{}{\bar G}.
\label{tinv}
\ee
Using this in \f{c1} we arrive at
\be
&&1=(1-D)G ^{-1} G_\s 
\nonumber\\
&&\!+\![1\!+\!(\Sigma_i  \!-\!\tilde V_i) G]^{-1}[(1\!+\!\Sigma_i G) D \!-\!\tilde V_i GD \!+\!\con{V}{}{\bar G} G]G^{-1} G_\s.    
\nonumber\\&&
\label{c2}
\ee
Now we choose the operator $D$ such that the last two terms cancel each other, i.e. 
\be
\con{V}{}{ \bar G} G\equiv \tilde V_i G D.
\label{D}
\ee
Using \f{1} in the form $G ^{-1} G_\s= (1+\Sigma_i G)^{-1}$ and subtracting
from \f{c2} the structure $1=(1-D) A^{-1} A+ B^{-1}B D$ we obtain finally
\ba
&(1-D) (1+\Sigma_i G)^{-1}\Sigma_i=
\nonumber\\
&[1\!+\!(\Sigma_i  \!-\!\tilde V) G]^{-1}[\tilde V_iGDG^{-1}\!-\!(1\!+\!\Sigma_i G)D(1\!+\!\Sigma_i G)^{-1}\Sigma_i].
\label{soven}
\end{align}
Together with the operator \f{D} and the effective potential $\tilde V_i$
defined by \f{veff} this is the desired generalized Soven equation. It is
written in operator form which becomes an algebraic equation in the
appropriate representation. In the operator form it is even valid in nonequilibrium and its time ordering can be treated e.g. in the framework of generalized Kadanoff and Baym formalism \cite{LSM97} with the help the Langreth/Wilkens rules \cite{LWI72}.  

Though introduced merely for mathematical
convenience, the operator $D$ corresponds to the concentration for impurity
scattering and the effective potential corresponds to the effective potential
in CPA. Therefore $D$ is called concentration operator hereafter. 

Let us illustrate this with the help of special cases. Choosing the averaged T-matrix approximation we have only one-particle functions, $K=G_\s$ and the closing by the concentration as a c-number $\bar G \to c$, such that we get from (\ref{veff}) $\tilde V_i=V$ which gives with (\ref{D}) $D=c$ such that the standard Soven equation \cite{VKE68} appears
\ba
(1-c)\,{\Sigma_i(p)\over 1+\Sigma_i(p) G(p)}
&=&c\, {V(p) -\Sigma_i(p)\over 1+[\Sigma_i(p)  -V(p)] G(p)}.
\label{soven1}
\nonumber\\
\end{align}

As a second, so far not known, example we give the explicit expressions for
the two-particle T-matrix. Then $K=G\cdot G_\s$ is a product in spatial
coordinates. Fourier-transform of the difference coordinates and gradient
expansion reveals then the structure [$p=({\bf p},\omega_p)$]
\ba
\left [1\!-\!D(p)\right ]{\Sigma_i(p)\over 1\!+\!\Sigma_i(p) G(p)}
&=& \!\!\!\!D(p){\tilde V_i(p) \!-\!\Sigma_i(p)\over 1\!+\![\Sigma_i(p)  \!-\!\tilde V_i(p)] G(p)}.
\nonumber\\
\label{soven2}
\end{align}
The concentration operator takes the form
\be
D(p)={\sum\limits_{\bar {\bf p},\bar \omega} V\left ({{\bf \bar p-p}\over 2}\right ) G
  (-\bar p)\over
  \tilde V_i(p)}
={\Sigma_H({\bf p})\over \tilde V_i(p)}
\label{d1}
\ee
noting the Hartree selfenergy $\Sigma_H$
and the effective potential reads
\ba
\tilde V_i(p)=
{\con{V}{KT_i}{\bar G}\over  \con{T}{_i}{\bar G} G_\s}
={\con{T}{_i}{\bar G}-\con{V}{}{\bar G}\over \con{T}{_i}{\bar G} G_\s}
={1\over G_\s(p)}-{\Sigma_H({\bf p})\over \Sigma_i(p) G(p)}.
\label{d2}
\end{align}
Here we have used (\ref{t}) for the first equality and (\ref{ide}) for the second one.
The channel T-matrix reads explicitly 
\ba
\Sigma_i(p)
=\!\! \sum\limits_{\bar p} T_i \left ({\bf {\bar
    p\!-\!p\over 2},{\bar p\!-\!p\over 2},p\!+\!\bar p},\omega_p+\omega_{\bar p}\right)G_\s(-\bar p).
\end{align}
It is instructive to see that introducing equations  \f{d1} and \f{d2} into \f{soven2} leads indeed to an identity. This is due to the fact that we have assumed that all quantities like T-matrix and selfenergy are known exactly. In case that we start with an approximation for the Green function or selfenergy we can use the above equation system and the generalized Soven form to iterate and to obtain approximations for the channel corrected propagators and selfenergies.

To summarize we have shown that the Dyson equation can be rewritten exactly into a form containing anomalous propagators and selfenergies if the repeated collisions in one channel are subtracted. The correction of the channel results into a selfenergy which obeys a generalized Soven equation irrespective of the actually form of approximation used for the T-matrix. We have derived this identity here which allows to improve any chosen approximation in the same way as the CPA does it with the averaged T-matrix approximation. The suggested procedure is to choose an approximation for the T-matrix and to calculate the self-consistent channel-corrected propagators by iteration of \f{d1} and \f{d2} and \f{soven2}.
Due to the versatile appearance of pairing and condensation phenomena ranging from nuclear,  to solid state up to plasma physics, the systematic improvement of any approximation used and the underlying equivalences of the different approaches might be helpful. The actual application of the proposed scheme is reserved for further work.

The discussions with Pavel Lipavsk\'y, Bretislav \v Sopik and Michael M\"annel
are gratefully mentioned.
This work was supported by DFG-CNPq project 444BRA-113/57/0-1 and the DAAD-PPP (BMBF) program. The financial support by the Brazilian Ministry of Science 
and Technology is acknowledged .

\bibliography{bose,kmsr,kmsr1,kmsr2,kmsr3,kmsr4,kmsr5,kmsr6,kmsr7,delay2,spin,spin1,refer,delay3,gdr,chaos,sem3,sem1,sem2,short,cauchy,genn,paradox,deform}

\end{document}